\UseRawInputEncoding 
\documentclass[12pt]{article}
\usepackage{latexsym}
\usepackage{graphicx}
\usepackage{dcolumn}
\usepackage{bm}
\usepackage{amsmath}
\usepackage{amssymb}
\usepackage{color}
\usepackage{slashed}
\usepackage{authblk}
\usepackage{hyperref} 
\title{Relativistic QFT from a Bohmian perspective: \\
A proof of concept}
\author{Hrvoje Nikoli\'c\footnote{e-mail: hnikolic@irb.hr}}
\affil{Theoretical Physics Division, Rudjer Bo\v{s}kovi\'{c} Institute, \\
P.O.B. 180, HR-10002 Zagreb, Croatia}

\begin{document}

\maketitle


\begin{abstract} 
Since Bohmian mechanics is explicitly nonlocal, it is widely believed that it is very hard,  
if not impossible, to make Bohmian mechanics compatible with relativistic quantum field theory (QFT).
I explain, in simple terms, that it is not hard at all to construct a Bohmian theory 
that lacks Lorentz covariance, but makes the same measurable predictions as relativistic QFT.
All one has to do is to construct a Bohmian theory 
that makes the same measurable predictions as QFT in {\em one} Lorentz frame,
because then standard relativistic QFT itself guarantees that those predictions are Lorentz invariant.
I first explain this in general terms, then I describe a simple Bohmian model that makes the same
measurable predictions as the Standard Model of elementary particles, after which I give some hints 
towards a more fundamental theory beyond Standard Model. 
Finally, I present a short story telling how my views of fundamental physics
in general, and of Bohmian mechanics in particular, evolved over time.      
\end{abstract}


\section{Introduction}

\begin{flushright}
{\it Basic prejudice: Bohmian mechanics is an interpretation of non-relativistic quantum mechanics, not of relativistic QFT.}
\end{flushright}

Bohmian mechanics (BM) is a possible solution of the problem of quantum measurement 
\cite{bohm,book-bohm,book-hol,book-durr,oriols,nikIBM}.
The most salient feature of BM  is that it is an {\em ontological} theory, i.e. a theory 
that explicitly describes ``real'' {\em ontic} stuff. For physicists who are not familiar with the meaning 
of that word, here is a brief explanation. For a given physical theory,
the ontic objects are those elements of the theory that directly correspond to actual physical objects 
that exist in Nature. In classical mechanics of one particle, for instance, a pointlike particle with a trajectory
${\bf X}(t)$ is ontic, while auxiliary quantities such as Lagrangian $L({\bf x},\dot{\bf x})$, Hamiltonian
$H({\bf x},{\bf p})$ or Hamilton-Jacobi function $S({\bf x},t)$ are not ontic. Those non-ontic 
quantities are just tools to compute the properties of the ontic objects. 
Likewise, in classical mechanics of $N$ particles, the ontic object is the set of $N$ trajectories
${\bf X}_1(t), \ldots, {\bf X}_N(t)$.
Another example is classical statistical
mechanics, which serves as an effective theory for a system of a large number $N$ of classical particles.
Thus the ontic object in classical statistical mechanics is again the $N$ trajectories,
while the probability distribution in the phase space 
is not ontic.\footnote{Probability is a theoretical tool for making predictions or explanations of the actual occurrences
(e.g.~the actual number of particles within a given region in 6-dimensional phase space)
when some relevant information about the actual system is unknown.
As such, probability is a non-ontic entity in both the Bayesian and the frequentist interpretation,
while the actual occurrences themselves are ontic.
The theoretical frequencies of occurrences computed by probability theory depend on our partial ignorance
about the actual system, while the actual frequencies do not depend on our ignorance. 
Failing to distinguish the former frequencies from the latter frequencies
is like failing to distinguish the map from the territory.
For a large $N$, however, owing to the law of large numbers, the two frequencies typically differ only by a negligible amount,
which resolves various conceptual confusions in statistical physics. For example,
this explains why the ``subjective'' Gibbs entropy, defined in terms of probabilities depending on our ignorance, in practice often can be treated as being the same as
``objective'' Gibbs entropy defined by actual frequencies.}
By analogy, in the Bohmian interpretation of non-relativistic 
quantum mechanics (QM), the ontic object is the set 
of $N$ trajectories, while the wave function $\psi({\bf x}_1,\ldots,{\bf x}_N,t)$ 
is best viewed as a law-like ``nomological'' object \cite{nomological}, 
akin to the classical Hamilton-Jacobi function $S({\bf x}_1,\ldots,{\bf x}_N,t)$. 
In BM, the quantum wave function $\psi$ determines velocities of quantum particles, in the same sense in which
the classical Hamilton-Jacobi function $S$ determines velocities of classical particles.
This should be distinguished from the standard interpretation of QM, which is not ontological because (i) it 
does not involve particle trajectories at all,
and (ii) because it does not offer a clear direct answer to the question 
whether $\psi$ is ontic or not.\footnote{In the standard interpretation of QM one often starts 
from the assumption that $\psi$ provides a {\em complete} description of the system. But 
$\psi$ itself is either ontic or non-ontic, and both options are conceptually problematic 
in a $\psi$-complete interpretation. The first option, that $\psi$ is ontic, is hard to reconcile with the view that
$\psi$ is just a probability amplitude that exhibits an apparent collapse upon measurement.
From this point of view, the second option of non-ontic $\psi$, 
in which case the ``collapse'' is nothing but an update of information,
seems much more sober. But if $\psi$ is non-ontic then something else should be ontic,
for otherwise one would need to abandon the view that Nature really exists out there, even when we don't observe it.  
And yet, if $\psi$ is both complete and non-ontic, then it is hard to understand what exactly {\em is} ontic.
This is why the standard interpretation cannot state clearly whether $\psi$ is ontic or not.  
A way out of this conundrum is to adopt a non-standard interpretation
in which $\psi$ is {\em not complete}, which indeed is the main idea behind the Bohmian interpretation.}  

Another important feature of BM is explicit nonlocality, in agreement with the Bell theorem \cite{bell,schol}
which states that any ontic formulation of QM (satisfying certain reasonable requirements \cite{tumulka}) 
must be nonlocal.  
For many physicists, however, nonlocality is hard to swallow. A typical nonlocality argument
against BM is something like this: Nonlocality implies that BM is not consistent with relativity,
which implies that it is not consistent with relativistic quantum field theory (QFT),
which implies that it is not consistent with the real world.
The goal of the present paper is to explain why is that argument wrong!
  
Since the argument against BM above is very simple, 
the counterargument should be simple too, for otherwise it would not convince many critics. 
But relativistic QFT is a technically complex theory\footnote{Philosophers of physics 
who find standard QFT textbooks technically formidable may learn a lot of QFT from \cite{teller}.}
\cite{bd2,ryder,chengli,schwartz}, 
and its Bohmian version is not less complex. 
Therefore a fully technical analysis would not be efficient.
For that reason, in this paper I present a simple conceptual non-technical analysis that,
hopefully, can easily be understood by a lot of readers. 
In other words, in this paper I only present a proof of concept.

The paper is organized as follows. In Sec.~\ref{SECgenid}, I present general ideas that explain how 
a Lorentz non-covariant theory can have Lorentz invariant measurable predictions. In particular,
I explain the general scheme how BM without Lorentz covariance can make measurable predictions
compatible with those of relativistic QFT. 
In Sec.~\ref{SECminimal}, I present a minimal Bohmian model that makes the same measurable predictions as 
the Standard Model of elementary particles, while in Sec.~\ref{SECmoref}, I give some hints 
towards a more fundamental theory beyond Standard Model.
Finally, in Sec.~\ref{SEClove}, I tell a personal story describing how my views of fundamental physics
in general, and of Bohmian mechanics in particular, evolved over time. 

To motivate and prepare the reader, 
each section (including this one) and subsection starts with a basic prejudice, that many readers may hold, 
but is going to be overcome, shaken up or at least questioned in that (sub)section.

\section{General ideas}
\label{SECgenid}

\begin{flushright}
{\it Basic prejudice: Bohmian QFT is hard.}
\end{flushright}

\subsection{A simple analogy: classical electromagnetism}
\label{SECanalogy}

\begin{flushright}
{\it Basic prejudice: Gauge potentials are not real.} 
\end{flushright}

In the standard interpretation of classical electromagnetism,
the ``real'' ontic object is the field 
\begin{equation}
F_{\mu\nu}=\partial_{\mu}A_{\nu}-\partial_{\nu}A_{\mu} . 
\end{equation}
The $F_{\mu\nu}$ is {\em measurable}, {\em gauge invariant} and {\em Lorentz covariant}. 
The gauge potential $A_{\mu}$, on the other hand, is not ontic, not measurable and not gauge invariant.

Such an interpretation, however, is not without conceptual problems. 
The mathematical formulation of the theory requires $A_{\mu}$, it cannot be eliminated from the theory \cite{wald}. 
Moreover, if $A_{\mu}$ is not ``real'', then it is hard to understand the Aharonov-Bohm effect \cite{AB}.

This motivates us to consider an alternative interpretation, according to  which
$A_{\mu}$ is a fundamental ``real'' ontic object.
Such an alternative interpretation has nontrivial consequences. 
For given $F_{\mu\nu}$, the potential $A_{\mu}$ is not unique. 
(This, of course, is a consequence of gauge invariance, because $A_{\mu}$ and $A'_{\mu}=A_{\mu}+\partial_{\mu}\lambda$ 
give the same  $F_{\mu\nu}$.) 
Hence $A_{\mu}$  can only be ontic if the gauge is completely {\em fixed}.
The simplest possibility of ontic $A_{\mu}=(\phi,{\bf A})$ is the Coulomb gauge
\begin{equation}
\bm{\nabla} \cdot {\bf A}=0, \;\;\;\; \bm{\nabla}^2\phi=-\rho ,
\end{equation}
which implies
\begin{equation}\label{phi_coul}
 \phi({\bf x},t)=\frac{1}{4\pi}\int d^3x' \, \frac{\rho({\bf x}',t)}{|{\bf x}-{\bf x}'|} .
\end{equation}
Hence this ontic theory is {\em not gauge invariant}, {\em not Lorentz covariant} 
and {\em not local} (nonlocality manifests as instantaneous action at a distance in (\ref{phi_coul})). 
And yet, this theory makes the same {\em measurable} predictions 
as the standard theory of electromagnetism. This is because, mathematically,
it {\em is} the standard theory, just written in a fixed gauge. 

This is a simple example of the proof of concept: 
An ontic theory which {\em violates} gauge invariance, 
Lorentz covariance and locality, yet all its measurable predictions 
{\em are} gauge invariant, Lorentz covariant and local.
Bohmian mechanics for relativistic QFT will turn out to be conceptually similar.

\subsection{Philosophical questions}

\begin{flushright}
{\it Basic prejudice: Philosophy is useless.} 
\end{flushright}

Now let me discuss some philosophical questions associated with the theory in Sec.~\ref{SECanalogy},
in which $A_{\mu}$ is interpreted as ``real'' ontic stuff.
The questions that follow may be hard for (most) physicists, yet easy for philosophers. 

First, what exactly is the {\em difference} between standard theory and this theory? 
The difference is that in this theory $A_{\mu}$ is an ontic object.
This, however, is a philosophical difference, so many physicists may have problems
to understand what that means. In this paper I will not attempt to explain the general meaning of the word {\em ontic}
any deeper than I explained it in Introduction (for more detailed explanations see e.g. \cite{norsen-book,durr-book2}).
Instead, I shall take the case of $A_{\mu}$ as an example that {\em illustrates} 
the utility and subtleness of the general notion of 
ontology, because, I believe, in this case its utility and subtleness is quite intuitive.
  
Second, those who understand what the notion of ontic means may still
ask the following question: What is the {\em point} of saying that $A_{\mu}$ is ontic? 
The point is mainly conceptual. Computations are the same as in the standard theory, but 
the notion of ontic $A_{\mu}$ 
helps to understand some aspects (e.g. Aharonov-Bohm effect) {\em intuitively}. 
The notion of ontology is mainly a thinking tool, not a computational tool. 
Those who prefer a shut-up-and-calculate approach to physics may not need this tool at all. 
 
Third, some mathematically intelligent readers may ask the following question:
Isn't this example mathematically {\em trivial}? 
Yes, it is trivial, but that is exactly the point! 
The point of the example is to realize that something which at first may seem {\em impossible} 
(local Lorentz covariant phenomena explained by nonlocal Lorentz non-covariant theory)
may in fact be very {\em easy}. 
This example is a proof of concept!

\subsection{General Lorentz (non)covariant theories, including QFT}
\label{SECgen}

\begin{flushright}
{\it Basic prejudice: A theory that makes Lorentz invariant measurable predictions must itself be Lorentz invariant.} 
\end{flushright}


After a warmup with a simple example in the previous subsections, now I deal
with the main subject of this work. In fact, the following subsection plays a central role
for the whole paper.  
Those who will fail to understand the following subsection will
probably not be able to understand the rest of the paper. To a large extent, the rest of the paper 
will be nothing but special cases and explicit realizations of general ideas introduced in the following subsection.
 
Relativistic QFT is ``local'' in the sense that its Lagrangian density is local
and that commutators between observables vanish outside the light-cone.
This, indeed, is the standard notion of ``locality'' in the QFT community,
but I shall refer to this kind of ``locality'' as S-locality, where S stands for 
``standard'', reminding us that it is associated with the standard (non-Bohmian) view of QFT.
The notion of S-locality should be distinguished from the notion of Bell locality (or the lack thereof) 
in the context of Bell theorem and BM, to which I refer simply as locality. 
The goal is to understand how measurable predictions of 
S-local Lorentz covariant QFT can be explained by a nonlocal
Lorentz non-covariant theory such as BM.
But I want the explanation to be simple, so I do not want to discuss 
technical details of QFT.
To achieve this goal, in this subsection I make the analysis very {\em abstract} and {\em general}.

Let me start with some abstract definitions. The definitions, however, will be informal 
(and hopefully intuitive), without any intention to make them mathematically precise.

For a physical theory $T$, let 
$P[T]$ be {\em the set of all practically measurable predictions implied by $T$}. 
Here ``practically measurable'' refers to measurements that can actually be 
performed in practice, by experimental techniques that exist in the presence or in the foreseeable future.
It does {\em not} refer to thought experiments that can only be performed in principle. 

Next, let $RQFT$ be relativistic QFT in the standard (non-Bohmian) formulation \cite{bd2,ryder,chengli,schwartz}.
I assume that the reader is already familiar with $RQFT$, so at this level I do not specify
any details. Finally, let 
$RQFT(S)$ denote $RQFT$ with calculations performed {\em in a Lorentz frame $S$}.
Again, I assume that the reader has already performed many $RQFT$ calculations
in a fixed Lorentz frame, so that I do not need to explain any technical details 
of $RQFT(S)$.

From standard QFT textbooks, it is known that the predictions of $RQFT$ are Lorentz invariant.
I write this fact formally as
\begin{equation}\label{lorinv}
 P[RQFT(S)]=P[RQFT(S')], \;\; \forall S,\forall S' .
\end{equation}
In words, the measurable predictions do not depend on the Lorentz frame in which
the calculations are performed.\footnote{QFT interactions are often studied 
in the interaction picture \cite{bd2} based on the operator $U(t)$ of unitary evolution, which satisfies
the Lorentz non-covariant equation $H_{\rm int}U=i\partial_tU$ with $H_{\rm int}$
being the interaction Hamiltonian. Nevertheless, after a Dyson expansion for $U(t)$,
one eventually gets Lorentz invariant Feynman rules for computing matrix elements of the scattering matrix.} 
Whatever the predictions are in the frame $S$, 
the predictions in any other frame $S'$ are the same.\footnote{For example, if Alice 
observes that her measuring apparatus shows 
that the energy of the particle is $E=7$ MeV, then Bob, who moves with respect 
to Alice with a relativistic velocity $v$, will also observe that {\em her} apparatus shows $E=7$ MeV.
The result that the Alice's apparatus shows $E=7$ MeV
can be obtained from calculations in {\em any} frame, this is why the result of measurement is Lorentz {\em invariant}
(and not just covariant). This invariant energy $E$ can be written as $E=-\eta_{\mu\nu}P^{\mu}U^{\nu}_{\rm appar}$,
where $P^{\mu}$ is the covariant 4-momentum of the particle, $U^{\mu}_{\rm appar}$ is the
covariant 4-velocity of the measuring apparatus and   
$\eta_{\mu\nu}$ is the Minkowski metric with signature $(-+++)$.}

As the final abstract informal definition, let $BM$ be ``Bohmian mechanics'', 
that is, some nonlocal ontic theory of a Bohmian type. 
At this point I do not need to specify much details of this theory.
I only assume that this theory is not Lorentz covariant, in the sense that
space and time in this theory are not treated on an equal footing. 
In the language of relativistic theories, one can say that BM is formulated 
in some fixed preferred Lorentz frame $S_0$.\footnote{More precisely, by ``fixed Lorentz frame''
in this context one means a fixed slicing of spacetime into space and time. The theory is still 
covariant under purely spatial rotations and, for that matter, under arbitrary time-independent
changes of spatial coordinates.}  

Now suppose that it is possible to construct a $BM$ theory such that
\begin{equation}\label{boxed}
 P[BM] \supseteq P[RQFT(S_0)] .
\end{equation}
In words, I suppose that $BM$ contains all the measurable predictions of $RQFT$ in the Lorentz frame $S_0$,
and, in addition, that $BM$ may contain some additional predictions on which $RQFT$ is agnostic. 
Then Lorentz invariance (\ref{lorinv}) implies
\begin{equation}
P[BM] \supseteq P[RQFT(S)], \;\; \forall S .
\end{equation}
In other words, if $BM$ reproduces the measurable predictions of $RQFT$ in the preferred
frame $S_0$ (assumption (\ref{boxed})), then it reproduces them in {\em all} frames $S$.
This is the general scheme explaining how Lorentz non-covariant theory can have Lorentz invariant predictions!

A nontrivial task is to construct a {\em concrete} theory with property (\ref{boxed}).
Actually many different constructions are possible, some of which I shall discuss in the next two sections.

\section{A minimal model}
\label{SECminimal}

\begin{flushright}
{\it Basic prejudice: All Bohmian models are models of particle trajectories.} 
\end{flushright}

\subsection{Motivation}

\begin{flushright}
{\it Basic prejudice: The goal of new scientific theories is to make new measurable predictions.} 
\end{flushright}

In general, we need a $BM$ theory that satisfies the assumption (\ref{boxed}), namely
$P[BM] \supseteq P[RQFT(S_0)]$.
A minimal Bohmian model is a theory for which 
\begin{equation}
P[BM] = P[RQFT(S_0)] .
\end{equation}
In words, a minimal Bohmian model has the same predictions as $RQFT$, 
without making any additional predictions.

As a proof of concept, I will present a conceptually simple model based on \cite{struyve-westman}.

\subsection{Fundamental Bohmian ontology} 

\begin{flushright}
{\it Basic prejudice: Matter is real.} 
\end{flushright}

I the simple model that I am now presenting the only fundamental ``real'' ontic stuff are bosonic fields, 
such as electromagnetic potentials ${\bf A}({\bf x},t)$ in the Coulomb gauge and 
Higgs field $h({\bf x},t)$. 
In this model, particle positions are {\em not} ontic. Fermionic fields are also {\em not} ontic,
which is related to the fact that fermionic fields do not have a well-defined classical limit.

It is instructive to compare this Bohmian bosonic {\em field} ontology with the more familiar
Bohmian {\em particle} ontology.
In Bohmian ontology for non-relativistic QM, a macroscopic object is made of a large number $N$ of 
particles, with positions ${\bf x}_1,\ldots ,{\bf x}_N$. 
The particle ontology is discrete.
By contrast, Bohmian ontology for QFT in this model is a set of bosonic fields, e.g. 
\begin{equation}
\phi({\bf x})=\{ {\bf A}({\bf x}), h({\bf x}), \ldots \} ,
\end{equation}
{\em at all space points} ${\bf x}$. 
The field ontology is continuous.
But at the macroscopic level both ontologies may look the same, because in both cases 
the macroscopic ontology is a pattern in space. The two ontologies differ at the fine-grained level,
but at the coarse-grained level they may look the same. From our everyday experience based on macroscopic 
phenomena, both ontologies seem viable. 

In principle, up to some technical complications (such as Gribov ambiguity), 
the field ontology can be defined for all bosonic fields in the Standard Model
(see e.g. \cite{struyve}). However, in a minimal model it is sufficient
to postulate that only the electromagnetic potential ${\bf A}({\bf x},t)$ is ontic \cite{struyve-westman}, 
while all other bosonic fields (Higgs, gluons, massive vector fields) are not ontic. 
I will explain it next.
(For a similar idea see also \cite{not_all}, where it is explained that it is not necessary that all 
particle species are ontic.)

Intuitively, we usually imagine that the material objects (such as tables and chairs) are made of atoms, 
namely of charged fermionic matter. How is that picture compatible with the idea that
fermionic matter is not ontic, while only electromagnetic potentials are ontic? 
Well, one reason why the material world looks ``real'' to us is because 
we {\em see} it, and seeing is nothing but observing light.
Since light is described by the electromagnetic potentials, the light is ontic.
This means that the pictures of material objects that we see are ontic, but the objects themselves are not.

What about other ways of experiencing the material objects, such as by touching them? 
Well, other senses can also be reduced to electromagnetic potentials, one way or another.
After all, all senses involve electric signals in the neural system, and electric signals 
can be encoded in the corresponding electromagnetic potentials.
Moreover, even if senses by living beings are ignored, a lot of information about charged matter 
can be encoded into the corresponding ${\bf A}({\bf x},t)$ fields. Thus the idea that
only ${\bf A}({\bf x},t)$ is ontic should not contradict our intuition and common sense.

\subsection{The mathematical construction}

\begin{flushright}
{\it Basic prejudice: The mathematics of relativistic QFT is more complicated than mathematics of non-relativistic QM.} 
\end{flushright}

Now I want to construct a Bohmian theory of bosonic fields in analogy with that for particles. 
The main idea is to write QFT in a form similar to non-relativistic QM,
in terms of wave functionals that satisfy a functional Schr\"odinger equation
(for details see e.g. \cite{hatfield}).
Here I briefly sketch the main ideas of the formalism, in a form that should be easy to understand 
at the intuitive conceptual level.

I start from bosonic field eigenstates $|\phi\rangle$, obeying $\hat{\phi}({\bf x})|\phi\rangle=\phi({\bf x})|\phi\rangle$. 
In terms of these eigenstates, all QFT states $|\Psi\rangle$ can be represented as functionals of bosonic fields 
\begin{equation}
 \Psi[\phi,t]=\langle\phi|\Psi(t)\rangle .
\end{equation}
The time evolution of the state is given by the functional Schr\"odinger equation 
\begin{equation}
 \hat{H}\Psi[\phi,t]=i\hbar\partial_t\Psi[\phi,t] ,
\end{equation}
where $\hat{H}$ is the QFT Hamiltonian derived from a Lorentz invariant Lagrangian.
In this way, relativistic QFT is written in a fixed Lorentz frame $S_0$. 
In this form, it does not look manifestly Lorentz covariant. Nevertheless, it is mathematically
equivalent to other, more covariant, formulations of QFT and the measurable predictions 
are Lorentz invariant \cite{hatfield}. 

The QFT Hamiltonian typically has the form
\begin{equation}\label{HQFT}
 \hat{H}=\int d^3x\, \frac{\hat{\pi}^2({\bf x})}{2}+V[\phi, \ldots] + \ldots ,
\end{equation}
where 
\begin{equation}
\hat{\pi}({\bf x})=-i\hbar\frac{\delta}{\delta\phi({\bf x})} 
\end{equation}
is the canonical momentum operator conjugate to $\phi({\bf x})$. The dots
$\ldots$ denote terms that do not depend on $\phi$ and $\hat{\pi}$, which refers to all fermionic fields 
and possibly those bosonic fields (if any) that are not interpreted as ontic.
For comparison, the Hamiltonian in non-relativistic QM typically has the form
\begin{equation}\label{Hpart}
 \hat{H}_{\rm NRQM}=\sum_{a=1}^N \frac{\hat{\bf p}_a^2}{2m_a}+V({\bf x}_1,\ldots,{\bf x}_N) ,
\end{equation}
where 
\begin{equation}
\hat{\bf p}_a=-i\hbar\frac{\partial}{\partial{\bf x}_a}.
\end{equation}
The associated Schr\"odinger equation in non-relativistic QM is 
\begin{equation}
\hat{H}_{\rm NRQM}\psi(x,t)=i\hbar\partial_t\psi(x,t) ,
\end{equation}
where $\psi(x,t)=\langle x|\psi(t)\rangle$ and $x=\{ {\bf x}_1,\ldots,{\bf x}_N\}$. 

We see that, essentially, the Hamiltonians (\ref{HQFT}) and (\ref{Hpart})
have the same mathematical form. This is especially clear if 
continuous QFT is regularized by a discrete lattice: $\int d^3x \rightarrow \sum_{\bf x}$. 
Therefore, Bohmian mechanics for QFT is a straightforward generalization of that for particles in 
non-relativistic QM. 
Essentially, Bohmian particle trajectories ${\bf X}_a(t)$ 
are replaced by Bohmian field trajectories $\Phi({\bf x},t)$.
More precisely, the equation for Bohmian particle trajectories   
\begin{equation}\label{bohmX}
\frac{d{\bf X}_a(t)}{dt} =  \left. \frac{-i\hbar}{2m_a} \,
\frac{\psi^{\dagger} \!\stackrel{\leftrightarrow\;}{ \mbox{\boldmath $\nabla$}_a }\!  \psi}
       {\psi^{\dagger}\psi} \right|_{x=X(t)} ,
\end{equation}
where $A \!\stackrel{\leftrightarrow\;}{ \mbox{\boldmath $\nabla$}_a }\! B 
\equiv A (\mbox{\boldmath $\nabla$}_a B) - (\mbox{\boldmath $\nabla$}_a A) B$,
is replaced by the equation for Bohmian field trajectories
\begin{equation}\label{bohmPhi}
 \frac{\partial{\Phi}({\bf x},t)}{\partial t} = \left.  \frac{-i\hbar}{2} \,
\frac{\Psi^{\dagger} {\frac{\!\stackrel{\leftrightarrow\;}{\delta}}{\delta\phi({\bf x})} }  \Psi}
       {\Psi^{\dagger}\Psi} \right|_{\phi=\Phi(t)} .
\end{equation}
The notations $x=X(t)$ and $\phi=\Phi(t)$ denote ${\bf x}_{a'}={\bf X}_{a'}(t),\,\forall a'$ 
and $\phi({\bf x}')=\Phi({\bf x}',t),\, \forall {\bf x}'$, respectively.
The daggers on $\psi^{\dagger}$ and $\Psi^{\dagger}$ denote complex conjugation plus transpose,
where transpose indicates that $\psi$ and $\Psi$ depend on additional degrees of freedom 
that are not written down explicitly. For $\psi$ those are spin indices (for particles with spin),
while for $\Psi$ those are non-ontic fields, namely all fermionic fields and possibly some bosonic ones. 
In expressions of the form $\psi^{\dagger}\cdots\psi$ and $\Psi^{\dagger}\cdots\Psi$,
all those additional degrees of freedom are summed over 
(e.g. $\psi^{\dagger}\psi\equiv \sum_s \psi^*_s\psi_s$, where $s=\{s_1,\ldots,s_N\}$ are spin indices of $N$ particles),
so the right-hand sides of (\ref{bohmX}) and (\ref{bohmPhi}) do not depend on those additional degrees.
More explicitly, $\Psi^{\dagger}\Psi$ in the denominator on the right-hand side of (\ref{bohmPhi}) 
is really a shorthand notation for
\begin{equation}\label{fermint}
 \Psi^{\dagger}\Psi = \Psi^{\dagger}[\phi,t]\Psi[\phi,t]=\int{\cal D}\chi \, 
\tilde{\Psi}^*[\phi,\chi,t]\tilde{\Psi}[\phi,\chi,t] ,
\end{equation}
where $\int{\cal D}\chi$ is the functional integral over all non-ontic fields 
(all fermionic and possibly some bosonic ones), and similarly for the numerator in (\ref{bohmPhi}).

\subsection{Measurable predictions}

\begin{flushright}
{\it Basic prejudice: Quantum theory is all about microscopic systems, not 
about macroscopic measuring apparatuses.} 
\end{flushright}

It is well known that the Bohmian version of non-relativistic QM makes the same measurable predictions 
as the standard (non-Bohmian) non-relativistic QM \cite{bohm,book-bohm,book-hol,book-durr,oriols}. 
I assume that the reader interested in a Bohmian version of relativistic QFT should already know
how that works for non-relativistic QM, so in this paper I will not discuss it explicitly.
Let me just say that the key is to study how the wave function of the measured particles
is entangled with wave function of the {\em measuring apparatus}. More details 
can be found e.g. in \cite{nikIBM}, or in sections on the quantum measurement theory in 
\cite{bohm,book-bohm,book-hol,book-durr,oriols}. 

The Bohmian theory of fields based on (\ref{bohmPhi}) reproduces the measurable predictions of standard QFT,
in the same way in which Bohmian theory of particles based on (\ref{bohmX}) reproduces the measurable predictions 
of standard non-relativistic QM.  
The theory based on (\ref{bohmPhi}) is defined in a preferred Lorentz frame,
so it reproduces the measurable predictions in this frame. But as I have explained in Sec.~\ref{SECgen},
the measurable predictions of standard $RQFT$ do not depend on the Lorentz frame, which implies that 
measurable predictions of the Bohmian theory based on (\ref{bohmPhi}) also do not depend on the Lorentz frame.

For an illustration, let me briefly comment the example of a two-slit experiment with a single photon.
In the Bohmian interpretation based on field ontology, there is no pointlike particle that travels only
through one slit. Instead, the ontic electromagnetic potential ${\bf A}({\bf x},t)$ 
of the photon, governed by (\ref{bohmPhi}), typically travels through both slits.\footnote{For some initial conditions
${\bf A}({\bf x},0)$ it may travel through one slit only, but for most initial conditions it travels through both slits.}
So why then do we observe only one localized particle? As for any other measurement, the key is to focus attention
on the behavior of the {\em macroscopic measuring apparatus}, 
not on the behavior of the microscopic measured object.\footnote{In BM with point-particle ontology, the  
two-slit experiment is often presented as an experiment where the Bohmian interpretation is particularly 
simple. However, this simplicity is misleading and leads to frequent misunderstandings of BM, sometimes 
even among Bohmian experts. With point-particle ontology it is very simple, indeed, to understand why does particle 
arrive at only one position at the detection screen. However, this position of a single particle is not what we 
really see in the experiment.
What we really see is a position of a {\em macroscopic pointer} associated with the measuring apparatus.
The position of the pointer is strongly correlated with the position of the particle, 
but the explanation of that correlation is not trivial.
The explanation requires a use of the theory of quantum measurements, which involves many particles.
Hence the explanation of the two-slit experiment in terms of a single particle arriving 
at a single position at the screen is an over-simplification that misses one of the essential ingredients 
of BM, namely the theory of quantum measurements. For a formulation of BM that attributes to macroscopic 
objects (including pointers of measuring apparatuses) 
a central role see \cite{nikIBM}.}
The fields $\Phi({\bf x},t)$ describe, among other things, the pattern of a macroscopic pointer associated with the detector.
It is this macroscopic pointer that shows only one definite measurement outcome.
The evolution equation (\ref{bohmPhi}) provides that probability of any given field configuration 
$\phi({\bf x})$ at any time $t$ is given by $\Psi^{\dagger}[\phi,t]\Psi[\phi,t]$, which is the same probability 
as that in the standard (non-Bohmian) QFT. 
Hence the probability of any definite outcome is the same as predicted by the standard QFT.

\section{Towards a more fundamental theory}
\label{SECmoref}

\begin{flushright}
{\it Basic prejudice: Relativistic QFT is fundamental.}
\end{flushright}

\subsection{Motivation}

\begin{flushright}
{\it Basic prejudice: The Standard Model of elementary particles is fundamental.} 
\end{flushright}

The model in Sec.~\ref{SECminimal} was minimal, in the sense that $P[BM] = P[RQFT(S_0)]$. 
This model assumed that the Standard Model of elementary particles \cite{chengli,schwartz} is fundamental.
However, it has one philosophically unappealing feature: 
its measurable predictions (and field action) {\em are} Lorentz invariant, while 
Bohmian equations are {\em not} Lorentz covariant. 
Of course, it is perfectly consistent for doing phenomenology, but philosophically it does not feel right 
for a {\em really fundamental} theory. 
The theory would be more elegant if both measurable predictions and Bohmian equations 
had the same symmetries. But due to nonlocality (which is a consequence of the Bell theorem), it is hard to make  
Bohmian equations Lorentz covariant. 
This suggests the possibility that, in the fundamental theory, the measurable predictions are also not Lorentz invariant. 
If so, then {\em Standard Model is not fundamental}.

From this point of view, BM gives a hint how to search for a more fundamental theory 
beyond the Standard Model \cite{nikIBM}. 
The main hint is that the fundamental action (beyond Standard Model) should not be Lorentz invariant. 
From a practical point of view, this means that measurable predictions (at very small distances at which
the Standard Model is not applicable) should not be Lorentz invariant.
Therefore we want a fundamental quantum theory $F$ such that
\begin{equation}\label{F1}
 P[F] \supset P[RQFT(S_0)] 
\end{equation}
and a corresponding Bohmian version of $F$ such that
\begin{equation}\label{F2}
 P[BM] = P[F] .
\end{equation}
It is hard to find a viable and convincing theory $F$ that satisfies (\ref{F1}).
But once $F$ is known, it is straightforward to construct the corresponding $BM$  that satisfies (\ref{F2}).  
Hence, in the rest of Sec.~\ref{SECmoref}, I will mostly deal with theories from the ``standard'' (non-Bohmian) 
point of view, without talking much about their Bohmian versions.  
 
Before attempting to propose an explicit theory $F$, one must have some general framework to work with.
In this section I argue that a promising general framework to find $F$ is  
the {\em condensed matter} framework. 
In this framework, the Standard Model of particle physics is just an 
{\em effective} theory,\footnote{For philosophers of physics not familiar 
with the idea of effective field theory I recommend \cite{wallace}.} 
in the same sense in which the field theory of phonons is an effective theory.\footnote{Of course, 
since phonons can be derived from the Standard Model, while the Standard Model, in this framework, 
can be derived from a more fundamental theory, 
one can say that the phonon theory is an effective theory of an effective theory, while 
the Standard Model is still more fundamental than the phonon theory. 
But from the Wilsonian point of view \cite{wilson,peskin,huang,gifted_amateur,shankar,tong,ydri}, 
different effective theories, such as the phonon theory and the Standard Model, 
are just different coarse grainings of the unknown fundamental theory.}
Such condensed-matter models are extensively discussed in the literature, 
for reviews see e.g. the books \cite{volovik,wen}. 
In this context it is also worth mentioning Bohmian models for fermionic particles based on the 
Dirac sea picture (see \cite{colin-struyve} and references therein), because the Dirac sea picture 
itself can be motivated by analogy with holes in condensed matter physics.

\subsection{Analogy between phonons and photons}

\begin{flushright}
{\it Basic prejudice: Photon is a true particle, phonon is a quasiparticle.} 
\end{flushright}

A phonon is a collective excitation of many particles, namely atoms. 
Hence, from an atomic point of view, phonon is a {\em quasiparticle}, 
not a ``true'' particle. Nevertheless, phonons are quite similar to particles 
of the Standard Model in general, and to photons in particular. 

What makes phonons similar to particles?
The key is to approximate the condensed-matter system of atoms by a collection of harmonic oscillators
\cite{kittel,simons,gifted_amateur}. 
Typically, in condensed matter physics one considers a crystal lattice with atom positions $x_a$, 
such that the potential energy of neighboring atoms is proportional to $(x_a-x_{a+1})^2$.
Thus each pair of neighboring atoms can be viewed as a harmonic oscillator.
But from elementary QM, we know that each harmonic oscillator has energy spectrum of the form
\begin{equation}
E_n=\hbar\omega \left(n+\frac{1}{2}\right),  
\end{equation}
where $n=0,1,2,3,\ldots$. This implies that
$n$ can be thought of as a number of ``quanta'', i.e. that 
$n$ behaves like a number of ``particles''.

A slightly more formal and precise way to describe this is as follows.
The harmonic oscillators decouple in some new collective coordinates (so-called normal modes), 
so let $k=1,\ldots ,N$ label $N$ decoupled harmonic oscillators. The Hamiltonian then takes the form
\begin{equation}
 \hat{H}=\sum_k \hbar\omega_k \left( \hat{n}_k +\frac{1}{2}\right) , 
\end{equation}
where
\begin{equation}
 \hat{n}_k =\hat{a}_k^{\dagger}\hat{a}_k, \;\;\;\; [\hat{a}_k,\hat{a}_{k'}^{\dagger}]=\delta_{kk'} .
\end{equation}
The complete set of Hamiltonian eigenstates is the infinite tower of states of the following form: 
\begin{itemize}
\item groundstate: $|0\rangle$, satisfies $\hat{a}_k|0\rangle=0$, 
\item 1-``particle'' states: $|k\rangle=\hat{a}_{k}^{\dagger}|0\rangle$, 
\item 2-``particle'' states: $|k_1,k_2\rangle=\hat{a}_{k_1}^{\dagger}\hat{a}_{k_2}^{\dagger}|0\rangle$, 
\item 3-``particle'' states: $\ldots$,
\end{itemize}
and so on. This formalism looks identical to QFT 
of elementary particles, such as  photons. 
Due to this analogy, the above quanta of lattice vibrations 
are called {\em phonons}. 
Therefore, formally, a phonon is not less a particle than a photon is a particle. 

Indeed, a photon is also a collective excitation. Just like a phonon is a collective excitation of atoms,
a photon is a collective excitation of the electromagnetic field. 
The electromagnetic field ${\bf E}({\bf x})$, ${\bf B}({\bf x})$ lives on a continuous 3-dimensional space, 
which can be thought of as a 3-dimensional lattice with spacing $l\rightarrow 0$.
But why then a photon is a ``true'' particle and phonon a ``quasiparticle''?
The difference is in the nature of the corresponding lattice {\em vertices}. 
For phonons, the vertices are themselves particles, namely atoms. 
Hence phonons emerge from atoms (not the other way around), 
so atoms are {\em more fundamental} particles than phonons. 
In this sense, a phonon is ``less'' a particle than an atom is a particle, 
so it makes sense to call phonon a ``quasiparticle''.

For photons, on the other hand, the ``vertices'' are simply fields ${\bf E}$, ${\bf B}$ at points ${\bf x}$. 
In the Standard Model (SM), the field vertices are not made of more fundamental particles.
This is different from condensed matter crystal vertices where the vertices are themselves particles. 
Hence, from the SM point of view, photon can be considered 
a fundamental particle, not ``quasiparticle''. 

However, the above arguments rest on the assumption that SM is fundamental.
If that assumption is wrong, i.e. if SM is not fundamental, then photon can be a quasiparticle
in exactly the same sense in which phonon is a quasiparticle.

\subsection{Emergence of Lorentz invariance and QFT}

\begin{flushright}
{\it Basic prejudice: Relativity is fundamental.} 
\end{flushright}

A phonon is a quantum of sound. In the long wavelength limit (that is, for wavelengths much larger than interatomic distances),
sound is described by a wave equation of the form
\begin{equation}\label{sound}
 \frac{1}{c_s^2} \frac{\partial^2\psi}{\partial t^2} - \mbox{\boldmath $\nabla$}^2 \psi =0 .
\end{equation}
It is the same wave equation as that for light, except that $c_s$ is the speed of sound, not the speed of light $c$.
In particular, (\ref{sound}) is Lorentz covariant with respect to Lorentz transformations defined with $c_s$ instead of $c$.
But (\ref{sound}) is derived from a non-relativistic theory of atomic motion, which itself is a more fundamental theory 
than the wave equation (\ref{sound}). This demonstrates that Lorentz covariance can be emergent from a more fundamental theory 
that itself is not Lorentz covariant. Lorentz covariance can be just an approximation valid at long distances.

Phonons, as quantum objects, can be described by (\ref{sound}) in which the classical field $\psi({\bf x},t)$
is promoted to an operator $\hat{\psi}({\bf x},t)$. In this way one obtains a QFT of phonons \cite{simons}. But this 
quantum theory is derived from non-relativistic QM of atoms. This demonstrates that Lorentz covariant QFT 
can be emergent from a more fundamental theory that itself obeys principles of non-relativistic QM.

All this suggests the following hypothesis. The Standard Model (SM) based on relativistic QFT 
is emergent from a more fundamental theory 
which itself obeys principles of non-relativistic QM. In other words, 
$RQFT$ is not a fundamental theory. The fundamental theory is non-relativistic QM describing some 
fundamental particles. Those fundamental particles are {\em not} photons, electrons, or other SM particles, 
because SM particles are really quasiparticles. The fundamental particles are new kinds of particles 
that might be discovered in future experiments, perhaps by colliders stronger than the currently existing ones 
such as Large Hadron Collider in CERN.  

At the moment, this hypothesis is just a general theoretical framework for a research program \cite{volovik,wen}. 
It is still far from a closed theory. But if this hypothesis is true, then Bohmian interpretation
of that theory is easy. It is just non-relativistic BM applied to those hypothetic fundamental particles.

\section[Dr. Bohm or: How I learned to stop worrying and love standard QM/QFT]{Dr. Bohm or: 
How I learned to stop worrying and love standard QM/QFT\protect\footnote{This title paraphrases 
the title of the Stanley Kubrick's film ``Dr. Strangelove or: How I learned to stop worrying and love the bomb''.
This Kubrick's title has been paraphrased in the titles of many physics papers, of which 
at least three \cite{love1,love2,love3} are relevant in the context of the present section.}}
\label{SEClove}

\begin{flushright}
{\it Basic prejudice: If you want to understand the most fundamental principles
of physics, study high-energy physics.} 
\end{flushright}

I am a ``Bohmian'', namely an adherent of a very non-standard interpretation of QM. 
So how can I not worry and love standard QM and QFT? As a ``Bohmian'', shouldn't I be strictly against standard QM/QFT?

No. If by standard QM one means instrumental QM (see e.g. \cite{peres}),
which is nothing but a set of rules for computing probabilities of measurement outcomes,  
then standard QM is fully compatible with Bohmian QM. 
I am not saying that they are equivalent; indeed, Bohmian QM offers answers to some questions on which instrumental 
QM has nothing to say. But I am saying that they are compatible, in the sense that no claim of instrumental QM contradicts 
any claim of Bohmian QM.

But still, if instrumental QM has nothing to say about certain questions, then why am I not worried? 
As a ``Bohmian'', I certainly do not consider those questions irrelevant. So how can I not worry about 
it when it says nothing about questions that I find relevant?

The answer is that I learned to stop worrying and love standard QM precisely because I know about Bohmian QM. 
But let me explain it from the beginning.

I always wanted to study the most fundamental aspects of physics. Consequently, as a student of physics, 
I was much more fascinated by topics such as particle physics and general relativity than about topics such 
as condensed matter physics. Therefore, my graduate study in physics and my PhD were in high-energy physics. 
Nevertheless, all the knowledge about QFT that I acquired as a high-energy physicist 
did not help me much to resolve one deep puzzle that really bothered me about QM. 
The thing that bothered me was how could Nature work like that? How could that possibly be? What could be a possible physical 
mechanism behind the abstract rules of QM? Should one conclude that there is no mechanism at all and that standard QM 
(including QFT) is the end of the story?

But then I learned about Bohmian QM, and that was a true revelation. It finally told me a possible story of how could that be. 
It didn't definitely tell how it {\em is} (there is no direct evidence that Bohmian mechanics is how Nature actually works), 
but it did tell how it {\em might} be. It is comforting to know that behind the abstract and seemingly paradoxical formalism of QM 
may lie a simple intuitive mechanism as provided by Bohmian QM. Even if this mechanism is not exactly how Nature really works, 
the simple fact that such a mechanism is possible is sufficient to stop worrying and start to love instrumental QM as a useful 
tool that somehow emerges from a more fundamental mechanism, even if all the details of that mechanism are not (yet) known.
It is somewhat like figuring out a mechanism by which a magician {\em could} pull out a rabbit from the hat,  
which makes one stop believing in true magic
without seeing how does the magician actually do it.

However, something important was still missing. Bohmian QM looks nice and simple for non-relativistic QM, 
but how about relativistic QFT? In principle, Bohmian ideas that I learned from the existing literature
worked also for relativistic QFT, 
but they did not look so nice and simple. My question was, can Bohmian ideas be modified such that it looks nice, 
simple and natural even for relativistic QFT? That question motivated my professional research on Bohmian QM/QFT 
and I published a lot of papers in which I was trying to formulate Bohmian mechanics in a Lorentz covariant
way (see e.g. \cite{nik-pilotQFT,nik-book} and references therein).

Nevertheless, I was not completely satisfied with my results. Even though I made several interesting modifications 
of Bohmian QM to incorporate relativistic QFT, neither of those modifications looked sufficiently simple and natural. 
Moreover, in \cite{nik2013}, a referee found a 
deep conceptual error that I was not able to fix. After that, I was no longer trying to modify Bohmian QM in 
a Lorentz covariant way.

However, a partial satisfaction came from a slightly different angle. In an attempt to make sense of 
an ``orthodox'' local 
non-ontological interpretation of QM, I developed a theory of solipsistic hidden variables \cite{niksolip},
which is a sort of a hybrid 
between Bohmian and Copenhagen QM. In this theory, an observer plays an important role, in the sense that 
Bohmian-like trajectories exist only for degrees of freedom of the observer and not for the observed objects. 
That theory helped me to learn that, in order to understand why do we observe what we observe, 
it is not necessary to know what exactly happens with observed objects. Instead, as solipsistic hidden variables demonstrate, 
in principle it can be understood even if the observed objects don't exist! It was a big conceptual revelation for me that 
shaped my further thinking about the subject.

But it does not mean that I became a solipsist. I certainly don't believe that observed objects don't exist. 
The important message of solipsistic hidden variables
is not that observed objects might not exist. The important message is that the exact nature of 
their existence is not really so important to explain their observation. That idea helped me a lot to stop worrying and 
love standard QM.

But that was not the end. As I said, in my younger days, my way of thinking was largely shaped by high-energy 
physics and not by condensed matter physics. I thought that condensed matter physics cannot teach me much about 
the most fundamental problems in physics. But it started to change in 2010, when, by accident, 
I saw in Feynman Lectures on Physics \cite{feyn3} that Bohmian mechanics is related to superconductivity. 
That suddenly made me interested in superconductivity. But superconductivity cannot be understood without understanding 
other more basic aspects of condensed matter physics, so gradually I became interested in condensed matter physics as a field. 
One very interesting thing about condensed matter physics is that it uses QFT formalism which is almost identical to QFT formalism 
in high-energy physics, but the underlying philosophy of QFT is very different. 
Condensed matter physics taught me to think about QFT in a different way than I was used to as a high-energy physicist.

One of the main conceptual differences between the two schools of thought on QFT is the interpretation of particle-like 
excitations resulting from canonical quantization of fields. 
In high-energy physics, such excitations are typically interpreted as elementary particles. 
In condensed-matter physics, they are usually interpreted as quasiparticles, such as phonons. 
Since I was also a Bohmian, that led me to a natural question: Does it make sense to introduce a Bohmian trajectory of a phonon? 
An obvious (but somewhat superficial) answer is that it doesn't make sense because only true particles, and not quasiparticles, 
are supposed to have Bohmian trajectories. But what is a ``true'' particle? What exactly does it mean that a photon is a 
``true'' particle and a phonon isn't?

It was this last question that led me to a crucial fundamental insight about Bohmian mechanics that shapes my current 
way of thinking.
As I discussed in \cite{nikIBM} and in the present paper, 
the analogy with condensed-matter quasiparticles such as phonons suggests a very natural resolution of the 
problem of Bohmian interpretation of relativistic QFT. According to this resolution, the so-called “elementary” 
particles such as photons and electrons described by relativistic QFT are not elementary at all. 
Instead, they are merely quasiparticles, just as phonons. Consequently, those relativistic particles 
do not have Bohmian trajectories at all. What does have Bohmian trajectories are some more fundamental 
particles described by non-relativistic QM. Non-relativistic QM (together with its Bohmian interpretation) is fundamental, 
while relativistic QFT is emergent. In this way, the problem of the Bohmian interpretation of relativistic QFT is circumvented 
in a very elegant way.

There is only one ``little'' problem with that idea. There is no experimental evidence that such more fundamental 
non-relativistic particles actually exist in Nature. Perhaps they will be discovered one day in the future, 
but at the moment it is only a theory. In fact, it is not even a proper theory, 
because it cannot tell anything more specific about the exact nature of those hypothetical non-relativistic particles.

Nevertheless, there are at least two good things about that hypothesis. First, unlike most other versions of Bohmian mechanics, 
this version makes a testable prediction. It predicts that, at very small distances not yet accessible to experimental 
technology, Nature is made of non-relativistic particles. Second, at distances accessible by current experimental technology, 
this version of Bohmian QM says that Bohmian trajectories are irrelevant. This means that, as far as relativistic QFT is 
concerned, I do not need to worry about Bohmian trajectories and can love standard QFT, without rejecting ``common sense'' 
in the form of non-relativistic Bohmian mechanics at some more fundamental scale. That's how I finally 
learned to stop worrying and love standard QM/QFT.

\section*{Acknowledgments}

The author is grateful to T. Juri\'c for useful comments on the manuscript.
This work was supported by the Ministry of Science of the Republic of Croatia.

\section*{Data availability statement}

My manuscript has no associated data.

\section*{Conflict of interest statement}

The author has no conflict of interest to declare that is relevant to the content of this article.

\end{document}